\documentclass[aps,twocolumn,epsfig,prb]{revtex4}

\usepackage{amsmath}
\usepackage{graphicx}

\def\nn{\nonumber}

\begin{document}

\title{Kohn Anomalies in Graphene Nanoribbons}

\author{Ken-ichi Sasaki}
\email[Email address: ]{SASAKI.Kenichi@nims.go.jp}
\affiliation{National Institute for Materials Science,
Namiki, Tsukuba 305-0044, Japan}

\author{Masayuki Yamamoto}
\affiliation{National Institute for Materials Science,
Namiki, Tsukuba 305-0044, Japan}

\author{Shuichi Murakami}
\affiliation{Department of Physics, Tokyo Institute of Technology,
Ookayama, Meguro-ku, Tokyo 152-8551, Japan}
\affiliation{PRESTO, Japan Science and Technology Agency,
Kawaguchi 332-0012, Japan}

\author{Riichiro Saito}
\affiliation{Department of Physics, Tohoku University,
Sendai 980-8578, Japan}

\author{Mildred S. Dresselhaus}
\affiliation{Department of Physics, Department of Electrical Engineering
and Computer Science, Massachusetts Institute of Technology, Cambridge,
MA 02139-4307}

\author{Kazuyuki Takai}
\affiliation{Department of Chemistry, Tokyo Institute of Technology,
Ookayama, Meguro-ku, Tokyo 152-8551, Japan}

\author{Takanori Mori}
\affiliation{Department of Chemistry, Tokyo Institute of Technology,
Ookayama, Meguro-ku, Tokyo 152-8551, Japan}

\author{Toshiaki Enoki}
\affiliation{Department of Chemistry, Tokyo Institute of Technology,
Ookayama, Meguro-ku, Tokyo 152-8551, Japan}

\author{Katsunori Wakabayashi}
\affiliation{International Center for Materials Nanoarchitectonics, 
National Institute for Materials Science,
Namiki, Tsukuba 305-0044, Japan}
\affiliation{PRESTO, Japan Science and Technology Agency,
Kawaguchi 332-0012, Japan}

\date{\today}
 
\begin{abstract}
 The quantum corrections to the energies 
 of the $\Gamma$ point optical phonon modes (Kohn anomalies)
 in graphene nanoribbons are investigated.
 We show theoretically that the longitudinal optical modes 
 undergo a Kohn anomaly effect, 
 while the transverse optical modes do not. 
 In relation to Raman spectroscopy, 
 we show that the longitudinal modes are not Raman active
 near the zigzag edge, while the transverse optical modes 
 are {\it not} Raman active near the armchair edge.
 These results are useful for identifying 
 the orientation of the edge of graphene nanoribbons
 by G band Raman spectroscopy, 
 as is demonstrated experimentally.
 The differences in the Kohn anomalies for nanoribbons and 
 for metallic single wall nanotubes are pointed out, and
 our results are explained in terms of pseudospin effects.
\end{abstract}

\maketitle

\section{Introduction}

Graphene nanoribbons (NRs) 
are rectangular sheets of graphene
with lengths up to several micrometers 
and widths as small as nanometers.~\cite{li08,jiao09,kosynkin09}
NRs can be regarded as unrolled single wall nanotubes (SWNTs).
Since SWNTs exhibit either metallic or semiconducting
behavior depending on the diameter and 
chirality of the hexagonal carbon lattice
of the tube,~\cite{saito92apl} 
it is expected that 
the electronic properties of NRs 
depend on the width and ``chirality'' of the
edge.~\cite{liu09,jia09,girit09}  
In fact it has been predicted that 
their electronic properties near the zigzag edge
are quite different from those near the armchair
edge.~\cite{kobayashi93,fujita96,nakada96,wakabayashi99} 
Thus the characterization of the NRs as well as SWNTs
is a matter of prime importance.

Raman spectroscopy has been widely used 
for the characterization of 
SWNTs~\cite{holden94,r642,sugano98,
w699,yu01,l818,bachilo02,strano03,doorn04,i1049}
and graphenes.~\cite{canifmmode04,ferrari06,yan07,pimenta07,casiraghi09,malard09}
Recently, it has been shown that
the frequencies and spectral widths 
of the $\Gamma$ point optical phonons 
(called the G band in Raman spectra)
depend on the position of the Fermi energy $E_{\rm F}$
and the chirality of the metallic
SWNT.~\cite{farhat07,nguyen07,wu07,das07} 
The Fermi energy dependence of the Raman spectra
can be used to determine the position of the Fermi energy,
and the chirality dependence of the Raman spectra
provides detailed information on the electronic 
properties near the Fermi energy of metallic SWNTs. 
These dependences 
originate from the fact that 
the conduction electrons of a metal partly 
screen the electronic field of the ionic lattice.
Kohn pointed out that 
the ability of the electrons to screen the ionic electric field
depends strongly on the geometry of the Fermi surface,
and this screening leads to a change 
in the frequency of a specific phonon 
and an increase in its dissipation
(the Kohn anomalies).~\cite{kohn59} 
The Kohn anomalies (KAs) in graphene systems are unique
in the sense that they can occur
with respect to the $\Gamma$ point phonons while
the KA in a normal metal occurs with respect to phonons
with $2k_{\rm F}$ where $k_{\rm F}$ is the Fermi wave vector.
The uniqueness of graphene comes from 
the geometry of the Fermi surface given by the Dirac
cone.~\cite{dubay02,piscanec04,lazzeri06prl,piscanec07,
caudal07,pisana07,ando08,sasaki08_curvat}

Since the geometry of the Fermi surface of NRs, and 
the energy band structure of NRs depend on the orientation of the
edge,~\cite{fujita96,nakada96,wakabayashi99} 
one may expect that 
the KAs of NRs depend on the ``chirality'' of the edge
like the KAs of metallic SWNTs.
In this paper, 
we study KAs in graphene NRs with zigzag and armchair edges.
A NR with a zigzag (armchair) edge
is hereafter referred to as a Z-NR (an A-NR) for simplicity
(see Fig.~\ref{fig:zigLOTO}(a) for an $N$ Z-NR with a width $W=N\ell$
where $\ell\equiv 3a_{\rm cc}/2$ and $a_{\rm cc}$ ($=0.142$nm) 
is the bond length).
We will show that the transverse optical (TO) phonon modes do not
undergo KAs in both A-NRs and Z-NRs.
The dissipation of the longitudinal optical (LO) phonon modes in Z-NRs
is suppressed as compared to those in A-NRs.
We also show that the LO (TO) modes 
are not Raman active in Z-NRs (A-NRs),
and that the KAs should be observed in only A-NRs.
This fact is useful in
identifying the orientation of the edge of NRs 
by G band Raman spectroscopy.

It is noted that the D band 
which consists of an intervalley K point phonon 
has been used to characterize 
the orientation of the edge of graphite and graphene.
Pimenta {\it et al}.~\cite{pimenta07} observed that
the intensity of the D band near the armchair edge 
of highly ordered pyrolytic graphite (HOPG)
is much stronger than that near the zigzag edge of HOPG.
This is confirmed for single layer graphene
by the experiments of You {\it et al}.~\cite{you08}
who also show that the D band 
has a very strong laser polarization dependence.
However, a strong D band intensity appears only $\sim$ 20 nm 
from the edges, so that the observation of this effect 
requires a precise experimental technique.~\cite{pimenta07}

This paper is organized as follows.
In Sec.~\ref{sec:ka},
we study KAs in Z-NRs and A-NRs.
In Sec.~\ref{sec:ri}, we point out that
the LO modes in Z-NRs (TO modes in A-NRs)
are not Raman active 
and show experimental results.
In Sec.~\ref{sec:ps}, 
the mechanism of the edge dependent 
KAs and Raman intensities
is explained in terms of the pseudospin.
A discussion and summary are given 
in Sec.~\ref{sec:dis} and Sec.~\ref{sec:sum}.

\section{Kohn anomalies}\label{sec:ka}

\subsection{Zigzag NRs}

\begin{figure}[htbp]
 \begin{center}
  \includegraphics[scale=0.4]{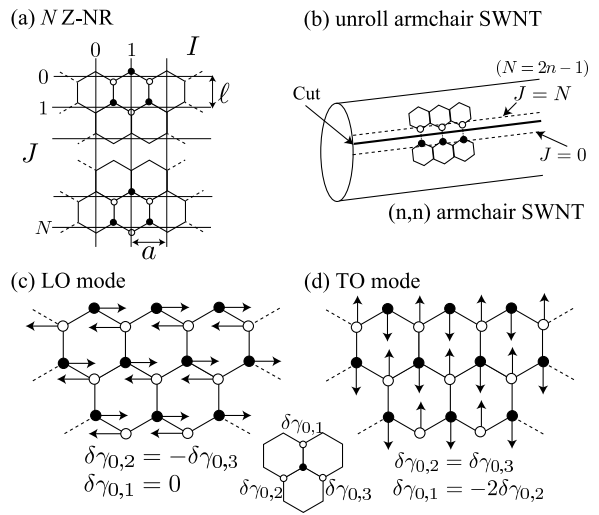}
 \end{center}
 \caption{
 (a) The lattice structure of a Z-NR.
 The solid (empty) circles denote the A (B) sublattice.
 We use integers $I$ $\in [0,M]$ and $J$ $\in [0,N]$ for the axes.
 The width (length) $W$ ($L$) of a Z-NR is given by 
 $N\ell$ ($aM$ where $a \equiv \sqrt{3}a_{\rm cc}$).
 (b) An $(n,n)$ armchair SWNT is cut along its axis and 
 flattened out to make an $N=2n-1$ Z-NR. 
 There is a metallic energy band in both structures 
 (see Fig.~\ref{fig:zigzag}(a) and (b)).
 (c,d) The phonon eigenvectors of the $\Gamma$ point 
 LO and TO modes are illustrated.
 The LO mode satisfies $\delta \gamma_{0,1}=0$ and
 $\delta \gamma_{0,2}  = - \delta \gamma_{0,3}$.
 The TO mode is characterized by 
 $\delta \gamma_{0,1} = -2 \delta \gamma_{0,2}$ and
 $\delta \gamma_{0,2}  = \delta \gamma_{0,3}$, and 
 $\delta \gamma_{0,1}$, $\delta \gamma_{0,2}$, $\delta \gamma_{0,3}$ are
 defined by the inset. 
 }
 \label{fig:zigLOTO}
\end{figure}

\begin{figure}[htbp]
 \begin{center}
  \includegraphics[scale=0.45]{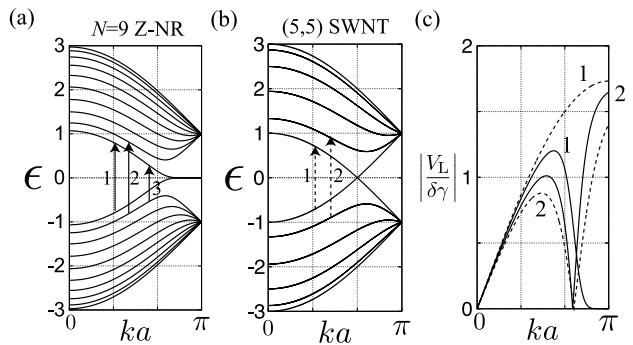}
 \end{center}
 \caption{
 The energy band structure of an $N=9$ Z-NR (a) and 
 that of a $(5,5)$ armchair SWNT (b).
 (c) $|V_{\rm L}/\delta \gamma|$ as a function of $ka$.
 The matrix elements of the vertical transitions, 
 denoted by the arrows (1,2) in (a) and (b), 
 are shown as the solid and dashed curves, respectively.
 }
 \label{fig:zigzag}
\end{figure}

KAs are relevant to 
the electron-phonon (el-ph) matrix element 
for electron-hole pair creation.
The electron-hole pair creation 
should be a vertical transition
for the $\Gamma$ point optical phonon,
and the KA effect, such as 
an increase in the dissipation of the phonon,
is possible only when the energy band gap of a NR is smaller 
than the energy of the phonon (about 0.2eV~\cite{saito98book}).
First, we study the energy band structures of Z-NRs
whose lattice and phonon modes are shown 
in Figs.~\ref{fig:zigLOTO}(a) and (c,d), respectively.
Z-NRs have a metallic energy band structure
regardless of their widths
as shown in Fig.~\ref{fig:zigzag}(a).
The metallicity of Z-NRs is due to 
the edge states forming a flat energy band 
at $E=0$.~\cite{fujita96}
Similarly, armchair SWNTs 
have a metallic energy band 
regardless of their diameters
as shown in Fig.~\ref{fig:zigzag}(b).~\cite{saito92apl} 
It is interesting to imagine that
an $N=2n-1$ Z-NR can be obtained from an $(n,n)$ armchair SWNT 
by cutting the circumferential C-C bonds along the tube axis
as shown in Fig.~\ref{fig:zigLOTO}(b).
For example, from a $(5,5)$ armchair SWNT, we get an $N=9$ Z-NR.
The metallicity of armchair SWNTs is preserved
in Z-NRs by the edge states.~\cite{sasaki06jpsj}

%%%%%%%%%%%%%%%%%%%%%%%%%%%%%
%%%%%%%%%%%%%%%%%%%%%%%%%%%%%

%%%%%%%%%%%%%%%%%%%%%%%%%%%%%
%%%%%%%%%%%%%%%%%%%%%%%%%%%%%

The phonon eigenvector of the LO (TO) mode 
is parallel (perpendicular) to the zigzag edge
as shown in Fig.~\ref{fig:zigLOTO}c(d).
By the displacement of a C-atom,
a bond length increases or decreases
depending on the position of the bond.
A change of bond length causes
a change of the three nearest-neighbor hopping integrals
from $-\gamma_0$ ($=2.7$eV)
to $-\gamma_0 + \delta \gamma_{0,a}$ ($a=1,2,3$)
(see the inset to Fig.~\ref{fig:zigLOTO}(c,d)).
The electron eigen function in the presence of $\delta \gamma_{0,a}$
is given by a linear combination of 
those in the absence of $\delta \gamma_{0,a}$.
In other words, 
a shift $\delta \gamma_{0,a}$ works as a perturbation 
by which an electron in the valence band 
may be transferred to a state in the conduction band.
This is an electron-hole pair creation process
due to a lattice deformation.
We derive the el-ph matrix element as follows.

Let $\psi_{\rm A}^{IJ}$ ($\psi_{\rm B}^{I'J}$) 
denote the wave function of an electron at an A-site (B-site) 
at site $IJ$ ($I'J$ where $I' \equiv I+1/2$).
Then $\sum_I (\psi_{\rm A}^{IJ})^* \delta \gamma_{0,3} \psi_{\rm B}^{I'J}$
is the amplitude for the process that 
an electron at the B-sites with $I'J$ is transferred 
by the perturbation $\delta \gamma_{0,3}$ 
into the A-sites with $IJ$
(see Fig.~\ref{fig:zigLOTO}(a)).
Note that $(\psi_{\rm A}^{IJ})^*$ indicates
complex conjugation of $\psi_{\rm A}^{IJ}$ 
and
$\delta \gamma_{0,3}$ for the LO and TO modes
is constant.
By introducing the Bloch function 
$(\phi_{\rm A}^{J},\phi_{\rm B}^{J})$ as 
$\psi_{\rm A}^{IJ}=(e^{iI(ka)}/\sqrt{M})\phi_{\rm A}^J$
and $\psi_{\rm B}^{I'J}=(e^{iI'(ka)}/\sqrt{M})\phi_{\rm B}^J$
where $k$ is the wave vector along the zigzag edge,
we obtain
$\sum_I (\psi_{\rm A}^{IJ})^* \delta \gamma_{0,3} \psi_{\rm B}^{I'J}=
(\phi_{\rm A}^{J})^* \delta \gamma_{0,3} e^{ika/2} \phi_{\rm B}^{J}$.
By deriving 
the matrix elements for $\delta \gamma_{0,2}$ and $\delta \gamma_{0,1}$
in a similar manner, we obtain 
the el-ph matrix element for 
a vertical electron-hole pair creation process 
such as $V+U$, where
\begin{align}
 \begin{split}
  & V= \sum_{J} 
  \begin{pmatrix}
   \phi^{J}_{\rm A} \cr 
   - \phi^{J}_{\rm B} 
  \end{pmatrix}^{\dagger}
  \begin{pmatrix}
   0 & e^{-i\frac{ka}{2}} \delta \gamma_{0,2} 
   + e^{i\frac{ka}{2}} \delta \gamma_{0,3} \cr 
   {\rm c.c.} & 0
  \end{pmatrix}
  \begin{pmatrix}
   \phi^{J}_{\rm A} \cr 
  \phi^{J}_{\rm B} 
  \end{pmatrix},
  \\
  & U=
  \sum_{J,J'} 
  \begin{pmatrix}
   \phi^{J'}_{\rm A} \cr 
   -\phi^{J'}_{\rm B} 
  \end{pmatrix}^{\dagger}
  \begin{pmatrix}
   0 & \delta \gamma_{0,1} \delta_{J',J+1} \cr
   \delta \gamma_{0,1} \delta_{J',J-1} & 0
  \end{pmatrix}
  \begin{pmatrix}
   \phi^{J}_{\rm A} \cr 
   \phi^{J}_{\rm B} 
  \end{pmatrix}.
 \end{split}
 \label{eq:m1}
\end{align}
$V$ and $U$ represent the el-ph interaction for 
the $\delta \gamma_{0,2}$ ($\delta \gamma_{0,3}$) perturbation 
which acts for the same $J$ 
and the $\delta \gamma_{0,1}$ perturbation 
which acts for the nearest neighbor pair of $J$ and $J'$, respectively.
For $V$ in Eq.~(\ref{eq:m1}),
${\rm c.c.}$ represents the complex conjugation of 
$e^{-i\frac{ka}{2}}\delta \gamma_{0,2}+e^{i\frac{ka}{2}}\delta \gamma_{0,3}$.
It is also noted that
the minus signs in front of $\phi^J_{\rm B}$ in Eq.~(\ref{eq:m1})
come from the fact that 
the Bloch function with energy $E$ in the conduction band 
is given by $(\phi^J_{\rm A},-\phi^J_{\rm B})$ when 
the Bloch function with energy $-E$ in the valence band 
is $(\phi^J_{\rm A},\phi^J_{\rm B})$.
This is a property of the nearest-neighbor tight-binding Hamiltonian 
with two sublattices.
The matrix elements $V$ and $U$ in Eq.~(\ref{eq:m1})
can be rewritten as 
\begin{align}
 V=V_{\rm T}+V_{\rm L}
\end{align}
where  
\begin{align}
\begin{split}
 & V_{\rm T} = 
 2i \left( \delta \gamma_{0,3} + \delta \gamma_{0,2} \right) 
 \cos \left(\frac{ka}{2}\right) 
 \sum_J {\rm Im}\left[ \phi^{J*}_{\rm A} \phi^J_{\rm B} \right], \\
 & V_{\rm L}= 2i \left( \delta \gamma_{0,3} -\delta \gamma_{0,2} \right) 
 \sin \left(\frac{ka}{2}\right) 
 \sum_J {\rm Re}\left[ \phi^{J*}_{\rm A} \phi^J_{\rm B} \right],
\end{split}
 \label{eq:a}
\end{align}
and 
\begin{align}
 U = 2i \delta \gamma_{0,1} \sum_{J}
 {\rm Im}\left[ \phi_{\rm A}^{J+1*}\phi_{\rm B}^J \right].
 \label{eq:U_zig}
\end{align}
By assuming that the perturbation
$\delta \gamma_{0,a}$ is proportional to a change of the bond length,
we have $\delta \gamma_{0,1} = 0$ and $\delta \gamma_{0,2}=-\delta \gamma_{0,3}$
for the LO mode, while
$\delta \gamma_{0,1} = -2 \delta \gamma_{0,2}$ 
and $\delta \gamma_{0,2}=\delta \gamma_{0,3}$
for the TO mode (see Fig.~\ref{fig:zigLOTO}(c) and (d)).

Thus, for the LO mode, both $V_{\rm T}$ and $U$ 
vanish because the LO mode satisfies 
$\delta \gamma_{0,2}+\delta \gamma_{0,3}=0$
and $\delta \gamma_{0,1}=0$, respectively.
The non-vanishing matrix element for the LO mode
is given by $V_{\rm L}$ only.
By setting $\delta \gamma_{0,2}=-\delta \gamma_{0,3}$
and introducing a shift $\delta \gamma$ due to a bond stretching
as $\delta \gamma_{0,3} \equiv \delta \gamma \cos(\pi/6)$ 
in Eq.~(\ref{eq:a}), we have
\begin{align}
 V_{\rm L}= 2\sqrt{3}i \delta \gamma
 \sin \left(\frac{ka}{2}\right) 
 \sum_J \phi^J_{\rm A} \phi^J_{\rm B}.
 \label{eq:Z_L}
\end{align}
From Eq.~(\ref{eq:Z_L}), it is understood that
the electron-hole pairs around $ka = 0$ 
are not excited since $|V_{\rm L}|$ 
is proportional to $\sin\left(ka/2\right)$.
Moreover, 
since the wave function of the the edge states
appears only on one of the two sublattice 
in the hexagonal unit cell,~\cite{fujita96} 
we have $\sum_J \phi_{\rm A}^J \phi_{\rm B}^J \approx 0$ 
in Eq.~(\ref{eq:Z_L}) for the edge states.
This suppresses the el-ph matrix element of 
electron-hole pair creation
for the edge states.
These facts can be checked by a numerical calculation
as shown in Fig.~\ref{fig:zigzag}(c) where we plot
$|V_{\rm L}/\delta \gamma|$ as a function of $ka$
by the solid curves for the two lowest energy 
electron-hole pair creation processes which are
denoted by the arrows in Fig.~\ref{fig:zigzag}(a).
In Fig.~\ref{fig:zigzag}(c), we see that
the edge states appearing as a flat energy band
at $2\pi/3 < ka < \pi$ 
do not contribute to electron-hole pair creation
(solid line 1).
In Fig.~\ref{fig:zigzag}(c),
we also plot $|V_{\rm L}/\delta \gamma|$
for a $(5,5)$ armchair SWNT for comparison
by the dashed curves for the two lowest energy 
electron-hole pair creation processes which are
denoted by the arrows in Fig.~\ref{fig:zigzag}(b).
As for the lowest energy electron-hole pairs
(the dashed line 1),
$|V_{\rm L}|$ increases with increasing $ka$ 
due to $\sin(ka/2)$ in Eq.~(\ref{eq:Z_L}).
This indicates a constant value of 
$\sum_J \phi_{\rm A}^J \phi_{\rm B}^J$ 
for the case of armchair SWNTs.
In Fig.~\ref{fig:zigzag}(c),
we see that the matrix element 
of the next lowest energy electron-hole pairs
vanishes at $k_0$ satisfying 
$\partial \epsilon/\partial k|_{k_0} = 0$ (van Hove singularity)
for both the Z-NR and armchair SWNT.
The same behavior is observed for higher sub-bands, and
a large density of states due to the van Hove singularities of the sub-bands 
is not effective in producing the electron-hole pair.

For the TO mode,
$V_{\rm L}$ vanishes owing to $\delta \gamma_{0,2}-\delta \gamma_{0,3}=0$.
Moreover, it can be shown that
\begin{align}
 {\rm Im}\left[ \phi^{J*}_{\rm A} \phi^J_{\rm B} \right]=0, \ \ 
 {\rm Im}[\phi_{\rm A}^{J+1*}\phi_{\rm B}^J]=0,
 \label{eq:imphi}
\end{align}
since an analytic solution of 
$(\phi_{\rm A}^J,\phi_{\rm B}^J)$ for Z-NRs is given 
in Ref.~\onlinecite{sasaki05prb} as
\begin{align}
 \begin{split}
  & \phi_{\rm A}^J = \left[ \frac{1}{g} \sin J \phi
  + \sin (J+1)\phi \right] C(g,\phi), \\
  & \phi_{\rm B}^J = \left[ \frac{\epsilon(g,\phi)}{g} 
  \sin (J+1) \phi \right] C(g,\phi).
 \end{split}
 \label{eq:wfz}
\end{align}
Here $g \equiv 2\cos(ka/2)$, and
$C(g,\phi)$ is a normalization constant, 
$\phi$ is the wave number 
which is determined by the boundary condition:
$\phi_{\rm A}^{N+1}=0$,
and $\epsilon(g,\phi)$ is the energy eigenvalue 
in units of $-\gamma_0$.
The energy dispersion relation is given by
$\epsilon(g,\phi)^2 = g^2 + 1 + 2g \cos \phi$.~\cite{sasaki05prb}
Since $C^*(g,\phi)C(g,\phi)$ is a real number, 
we get Eq.~(\ref{eq:imphi}).
As a consequence, we have $V_{\rm T}=0$ and $U=0$
in Eqs.~(\ref{eq:a}) and (\ref{eq:U_zig}).
Thus, the TO modes give both $V=0$ and $U=0$
for any vertical electron-hole pair creation process, 
and the TO modes decouple from the electron-hole pairs.
This shows the absence of the KA for the TO modes 
in Z-NRs.

A renormalized phonon energy is written 
as a sum of the unrenormalized energy $\hbar \omega_0$
and the self-energy $\Pi(\omega_0)$.
Throughout this paper, 
we assume a constant value for $\hbar \omega_0$
as $\hbar \omega_0 = 1600 {\rm cm}^{-1}$ 
(0.2eV) both for the LO and TO modes.
The self-energy is given by
time-dependent second-order perturbation theory as
\begin{align}
 \Pi(\omega) = &
 2 \sum_{\rm eh} \left(
 \frac{|V_{\rm L}|^2}{\hbar \omega -E_{\rm eh}+i\Gamma/2}
 - \frac{|V_{\rm L}|^2}{\hbar \omega +E_{\rm eh}+i\Gamma/2}
 \right) \nn \\
 & \times \left(f_{\rm h}-f_{\rm e}\right),
 \label{eq:PI}
\end{align}
where the factor 2 comes from spin degeneracy,
$f_{\rm h,e}=(1+\exp(\beta(E_{\rm h,e}-E_{\rm F}))^{-1}$
is the Fermi distribution function,
$E_{\rm e}$ ($E_{\rm h}$) is the energy of an electron (a hole),
and $E_{\rm eh}\equiv E_{\rm e}-E_{\rm h}$
is the energy of an electron-hole pair. 
In Eq.~(\ref{eq:PI}), 
the decay width $\Gamma$ is determined self-consistently
by $\Gamma/2= - {\rm Im} \left[ \Pi(\omega) \right]$.
The self-consistent calculation begins by putting
$\Gamma/2=\gamma_0$ into the right-hand side of Eq.~(\ref{eq:PI}).
By summing the right-hand side, we have a new $\Gamma/2$ via
$\Gamma/2= - {\rm Im} \left[ \Pi(\omega) \right]$ and we put it 
into the right-hand side again.
This calculation is repeated until $\Pi(\omega)$ is converged.
We use Eq.~(\ref{eq:Z_L})
with $\delta \gamma \equiv g_{\rm off}u(\omega)/\ell$
for $V_{\rm L}$ in Eq.~(\ref{eq:PI}).
Here $g_{\rm off}$ is the off-site electron-phonon matrix element 
and $u(\omega)$ is the amplitude of the LO mode.
We adopt $g_{\rm off} = 6.4$ eV.~\cite{jiang05prb} 
A similar value is obtained by a first-principles calculation 
with the local density approximation.~\cite{porezag95}
We use a harmonic oscillator model which gives 
$u(\omega)=\sqrt{\hbar/2M_c N_u \omega}$ where
$M_c$ is the mass of a carbon atom and $N_u$ 
is the number of hexagonal unit cells.

\begin{figure}[htbp]
 \begin{center}
  \includegraphics[scale=0.6]{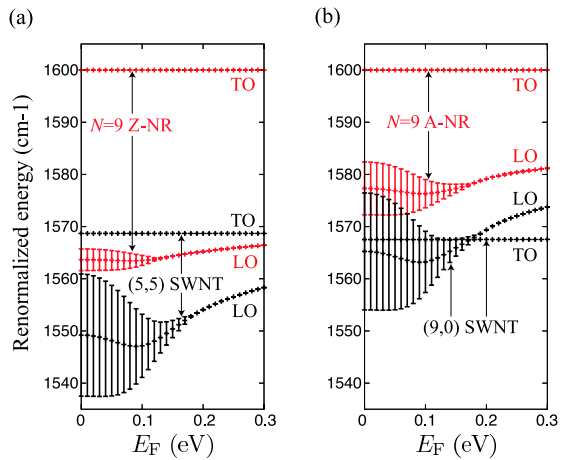}
 \end{center}
 \caption{(color online)
 (a) The $E_{\rm F}$ dependence of the renormalized energies of 
 the LO and TO modes in a $N=9$ Z-NR (red) 
 and of the LO and TO modes in a $(5,5)$ SWNT (black).
 (b) The $E_{\rm F}$ dependence of the renormalized frequencies of 
 the LO and TO modes in a $N=9$ A-NR (red) 
 and of the LO and TO modes in a $(9,0)$ SWNT (black).
 }
 \label{fig:EF}
\end{figure}

In Fig.~\ref{fig:EF}(a), 
we plot the renormalized energy
$\hbar \omega_0 + {\rm Re} \left[\Pi(\omega_0) \right]$
as a function of $E_{\rm F}$ 
for the LO and TO modes in a $N=9$ Z-NR 
at room temperature (300K).
Since the TO mode decouples from electron-hole pairs,
the self-energy $\Pi(\omega_0)$ vanishes and
the frequency of the TO mode 
does not change from $\omega_0$.
On the other hand, the LO mode exhibits a KA effect.
The error-bars extending up to $\pm \Gamma/2$ 
in Fig.~\ref{fig:EF} represent the broadening
of the phonon frequency due to the finite life time
of the phonon.
The LO mode shows the largest broadening effect of 
$\Gamma \approx 5$cm$^{-1}$ when $E_{\rm F}=0$eV.
The value of $\Gamma$ decreases quickly 
when $E_{\rm F} > 0.1$eV.
This is because of the Pauli exclusion principle 
by which a resonant decay of the LO mode is forbidden
when $E_{\rm F} > \hbar \omega_0/2$.~\cite{sasaki08_curvat}

For comparison,
we show the renormalized energies of the LO and TO modes
for a $(5,5)$ armchair SWNT as a function of $E_{\rm F}$ 
by the black curves in Fig.~\ref{fig:EF}(a).
%The TO mode does not undergo the KA effect
%even in a $(5,5)$ SWNT.
The TO mode exhibits no broadening but a softening 
with a constant energy $\sim -30$ cm$^{-1}$.
The absence of broadening is due to the fact that 
the Bloch function can be taken as a real number 
for lowest energy sub-bands, i.e.,  
for vertical transition denoted by the dashed line 1 
in Fig.~\ref{fig:zigzag}(b)
even when a Z-NR is rolled to form an armchair SWNT. 
The details are given in Sec.~\ref{sec:ps}.
For the LO mode, 
the broadening of the Z-NR
is smaller than that of the armchair SWNT
because $|V_{\rm L}|$ of a Z-NR
is smaller than that of armchair SWNTs
for the lower energy bands (see Fig.~\ref{fig:zigzag}(c)).
In fact, 
the real part of the right-hand side of Eq.~(\ref{eq:PI})
is a negative (positive)
value when $E_{\rm eh}> \hbar \omega_0$ 
($E_{\rm eh}< \hbar \omega_0$).
Thus, electron-hole pairs with higher (lower) energy
contribute to the frequency softening
(hardening).~\cite{sasaki08_curvat,sasaki08_chiral} 
Since the energies of the edge states
for Z-NR
are smaller than $\hbar \omega_0$,
the edge states may contribute a frequency hardening 
like the one shown around $E_{\rm F}=0$ for a $(5,5)$ SWNT.
The absence of the hardening confirms that 
the edge states do not contribute to $\Pi(\omega_0)$
because $|V_{\rm L}|$ is negligible.

%%%%%%%%%%%%%%%%%%%%%%%%%%%%%
%%%%%%%%%%%%%%%%%%%%%%%%%%%%%

It is noted that, 
for the renormalized energies of the LO and TO modes in NRs shown in Fig.~\ref{fig:EF}, 
we have not included all the possible intermediate electron-hole pair
states created by a given phonon mode in evaluating $\sum_{\rm eh}$ in
Eq.~(\ref{eq:PI}). 
For example, vertical transition denoted by the arrow 3
in Fig.~\ref{fig:zigzag}(a) may be included 
as a possible intermediate state in evaluating $\sum_{\rm eh}$ 
in Eq.~(\ref{eq:PI})
although such intermediate state 
does not satisfy the momentum conservation. 
%The renormalized energies of the LO and TO modes in NRs
%including all the possible intermediate electron-hole pair
%states are shown in Appendix~\ref{app:full}. 
In this paper, we do not consider the contribution of 
momentum non-conserving electron-hole pair creation processes 
in evaluating $\sum_{\rm eh}$ in Eq.~(\ref{eq:PI}). 
%This is an approximation which works well for thin NRs.

\subsection{Armchair NRs}

Next we study the KA in A-NRs.
The zigzag SWNTs are cut along their axis and flattened out
to make A-NRs.
It is known that one third of zigzag SWNTs 
exhibit a metallic band structure.~\cite{saito92apl}
It is interesting to note that
if we cut the bonds along the axis of a metallic zigzag SWNT 
in order to make an A-NR,
the obtained A-NR has an energy gap. 
Namely, unrolling a metallic $(3i,0)$ SWNT 
results in a $N=3i-1$ A-NR with an energy gap. 
Instead, unrolling a semiconducting $(3i+1,0)$ SWNT 
results in a gap-less $N=3i$ A-NR and
unrolling a semiconducting $(3i+2,0)$ SWNT 
results in a $N=3i+1$ A-NR with an energy gap.
The one-third periodicity of metallicity is maintained 
even if zigzag SWNTs are unrolled by cutting the bonds.
Since metallicity is a necessary condition for the KA,
we study the KA in $N=3i$ metallic A-NRs here.

\begin{figure}[htbp]
 \begin{center}
  \includegraphics[scale=0.4]{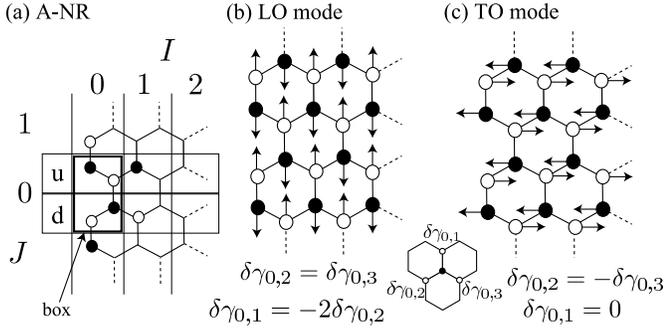}
 \end{center}
 \caption{ 
 (a) The lattice structure of an A-NR.
 (b,c)
 The eigenvectors of the $\Gamma$ point 
 LO and TO phonon modes are illustrated.
 The LO mode satisfies $\delta \gamma_{0,1} = -2\delta \gamma_{0,2}$ and
 $\delta \gamma_{0,2}  = \delta \gamma_{0,3}$.
 The TO mode is characterized by $\delta \gamma_{0,1} = 0$ and
 $\delta \gamma_{0,2}  = -\delta \gamma_{0,3}$.
 }
 \label{fig:armLOTO}
\end{figure}

In order to specify the lattice structure of an A-NR,
we use integers $I$ $\in [0,N]$ and $J$ $\in [0,M]$
in Fig.~\ref{fig:armLOTO}(a).
In a box specified by $IJ$ in Fig.~\ref{fig:armLOTO}(a),
there are two A atoms and two B atoms.
For convenience,
we divide the two A (B) atoms into 
up-A (up-B) and down-A (down-B),
as shown in Fig.~\ref{fig:armLOTO}(a).
The wave function then has four components:
$(e^{i(k2\ell)J}/\sqrt{M})
(\phi^I_{\rm uA},\phi^I_{\rm uB},\phi^I_{\rm dA},\phi^I_{\rm dB})^t$
where $k$ is the wave vector along the armchair edge.

%%%%%%%%%%%%%%%%%%%%%%%%%%%%%
%%%%%%%%%%%%%%%%%%%%%%%%%%%%%

In Fig.~\ref{fig:armLOTO}(b) and (c), 
we show phonon eigenvectors
of the $\Gamma$ point LO and TO modes, respectively.
The eigenvector of the LO (TO) mode 
is parallel (perpendicular) to the armchair edge.
The LO mode satisfies
$\delta \gamma_{0,1} = -2\delta \gamma_{0,2}$ 
and $\delta \gamma_{0,2}=\delta \gamma_{0,3}$, 
while the TO mode satisfies
$\delta \gamma_{0,1} = 0$ and $\delta \gamma_{0,2}=-\delta \gamma_{0,3}$.
Since $\delta \gamma_{0,2}$ and $\delta \gamma_{0,3}$
are perturbations that do not mix $\phi_{\rm u}$ and $\phi_{\rm d}$,
the electron-hole pair creation matrix element 
can be divided into the following two parts:
\begin{align}
\begin{split}
 & V_{\rm u} = \sum_{I,I'} 
 \begin{pmatrix}
  \phi^{I'}_{\rm uA} \cr
  -\phi^{I'}_{\rm uB}
 \end{pmatrix}^{\dagger}
 \begin{pmatrix}
  0 & \delta \gamma_{0,3} \delta_{I',I} + \delta \gamma_{0,2} \delta_{I',I+1} \cr
  {\rm h.c.} & 0 \cr
 \end{pmatrix}
 \begin{pmatrix}
  \phi^I_{\rm uA} \cr
  \phi^I_{\rm uB} 
 \end{pmatrix},
 \\
 &
 V_{\rm d} = \sum_{I,I'} 
 \begin{pmatrix}
  \phi^{I'}_{\rm dA} \cr
  -\phi^{I'}_{\rm dB}
 \end{pmatrix}^{\dagger}
 \begin{pmatrix}
  0 & \delta \gamma_{0,2} \delta_{I',I} + \delta \gamma_{0,3} \delta_{I',I-1} \cr
  {\rm h.c.} & 0
 \end{pmatrix}
 \begin{pmatrix}
  \phi^I_{\rm dA} \cr
  \phi^I_{\rm dB}
 \end{pmatrix},
\end{split}
 \label{eq:m2}
\end{align}
where the Hermite conjugate (${\rm h.c.}$) of $V_{\rm u}$ ($V_{\rm d}$)
is defined as 
$\delta \gamma_{0,3} \delta_{I',I} + \delta \gamma_{0,2} \delta_{I',I-1}$
($\delta \gamma_{0,2} \delta_{I',I} + \delta \gamma_{0,3} \delta_{I',I+1}$).
We can rewrite Eq.~(\ref{eq:m2}) as
\begin{align}
\begin{split}
 &
 V_{\rm u} = 2i \sum_{I} \left(
 \delta \gamma_{0,3} {\rm Im} \left[ \phi_{\rm uA}^{I*} \phi_{\rm uB}^I \right]
 + \delta \gamma_{0,2} {\rm Im} \left[\phi_{\rm uA}^{I+1*} \phi_{\rm uB}^I
 \right]
 \right), 
 \\
 & 
 V_{\rm d} = 2i \sum_{I} \left(
 \delta \gamma_{0,2} {\rm Im} \left[\phi_{\rm dA}^{I*} \phi_{\rm dB}^I \right]
 + \delta \gamma_{0,3} {\rm Im} \left[
 \phi_{\rm dA}^{I*} \phi_{\rm dB}^{I+1} \right]
 \right).
\end{split}
 \label{eq:Vud}
\end{align}
The perturbation $\delta \gamma_{0,1}$
mixes $\phi^I_{\rm u}$ and $\phi^I_{\rm d}$ as
\begin{small}
\begin{align}
 U = \sum_{I}
 \begin{pmatrix}
  \phi^{I}_{\rm uA} \cr
  -\phi^{I}_{\rm uB} \cr
  \phi^{I}_{\rm dA} \cr
  -\phi^{I}_{\rm dB}
 \end{pmatrix}^\dagger
 \begin{pmatrix}
  0 & 0 & 0 & e^{ik2\ell} \delta \gamma_{0,1} \cr
  0 & 0 & \delta \gamma_{0,1} & 0 \cr
  0 & \delta \gamma_{0,1} & 0 & 0 \cr
  e^{-ik2\ell} \delta \gamma_{0,1} & 0 & 0 & 0
 \end{pmatrix}
 \begin{pmatrix}
  \phi^I_{\rm uA} \cr
  \phi^I_{\rm uB} \cr
  \phi^I_{\rm dA} \cr
  \phi^I_{\rm dB}
 \end{pmatrix},
 \label{eq:U_arm}
\end{align}
\end{small}
so that $U$ can be rewritten as
\begin{align}
 U = i 2 \delta \gamma_{0,1}
 \sum_I \left( {\rm Im}
 \left[e^{ik2\ell} \phi^{I*}_{\rm uA} \phi^I_{\rm dB}\right]
 - {\rm Im}
 \left[ \phi^{I*}_{\rm uB} \phi^I_{\rm dA}\right]
 \right).
 \label{eq:A_L}
\end{align}

Now, it can be shown that each energy eigenstate
satisfies the following equations (see Appendix~\ref{app:mirror}),
\begin{align}
 \begin{split}
  & \sum_I \phi_{\rm uA}^{I*} \phi_{\rm uB}^I
  =\sum_I \phi_{\rm dA}^{I*} \phi_{\rm dB}^I,
  \\
  & \sum_I \phi_{\rm uA}^{I+1*} \phi_{\rm uB}^I
  =\sum_I \phi_{\rm dA}^{I*} \phi_{\rm dB}^{I+1}.
 \end{split}
 \label{eq:cor}
\end{align}
Due to these conditions,
the TO mode causes a special cancellation between
$V_{\rm u}$ and $V_{\rm d}$ as $V_{\rm u}+V_{\rm d}=0$
since $\delta \gamma_{0,2}+\delta \gamma_{0,3}=0$.
In addition, we obtain $U=0$ from $\delta \gamma_{0,1}=0$.
Thus the TO modes in A-NRs decouple from electron-hole pairs 
and do not undergo a KA.
For the LO mode, on the other hand, 
there is no cancellation between $V_{\rm u}$ and $V_{\rm d}$,
and the matrix element for the LO mode is given by 
$V_{\rm L} \equiv U+V_{\rm u}+V_{\rm d}$.
By setting $\delta \gamma_{0,2} = \delta \gamma_{0,3}$, 
$-2 \delta \gamma_{0,3} = \delta \gamma$ and 
$\delta \gamma_{0,1} = \delta \gamma$, 
we calculate $V_{\rm L}$ and put it into Eq.~(\ref{eq:PI})
to obtain $\Pi(\omega_0)$.

The solid curves in Fig.~\ref{fig:EF}(b) 
give the $E_{\rm F}$ dependence of the renormalized frequencies 
$\hbar \omega_0 + {\rm Re}\left[ \Pi(\omega_0)\right]$
for the LO and TO modes in a $N=9$ A-NR at room temperature.
The frequency of the TO mode is given by $\omega_0$ showing that
the TO mode decouples from electron-hole pairs.
The LO mode undergoes a KA.
For comparison, we show the renormalized frequency 
of the LO and TO modes in a $(9,0)$ zigzag SWNT as the 
black curves in Fig.~\ref{fig:EF}(b).
It is found that $|\Pi(\omega_0)|$ 
for the LO mode of a $N=9$ A-NR is 
smaller than $|\Pi(\omega_0)|$ 
for the LO mode of a $(9,0)$ zigzag SWNT.
It is because 
there are two linear energy bands near the K and K' points
in metallic zigzag SWNTs, 
while there is only one linear energy band in metallic A-NRs,
and the KAs in A-NRs are suppressed slightly
as compared to those in metallic zigzag SWNTs.
We note that the broadening in A-NRs is still larger than 
that in Z-NRs because of the absence of the edge states
near the armchair edges.
The TO mode of a $(9,0)$ zigzag SWNT is down shifted. 
However, there is no dependence on $E_{\rm F}$, which indicates that
only high energy electron-hole pairs
contribute to the self-energy.
We will explore the KA effect for the TO mode in zigzag SWNTs
in Sec.~\ref{sec:ps}.

We have considered NRs with a long length ($10\mu m$)
in calculating the self-energies $\Pi(\omega_0)$ shown in 
Fig.~\ref{fig:EF}(a) and (b).
For NRs with short lengths,
the effect of the level spacing on $\Pi(\omega)$ is not negligible.
For example,
the level spacing in armchair SWNTs 
becomes $0.12$eV when $L=30$nm, which is comparable to 
$\hbar \omega_0/2$.
%since the Fermi velocity $v_{\rm F}$ 
%satisfies $2\pi \hbar v_{\rm F}\approx 3.6[{\rm eV}\cdot{\rm nm}]$.
Thus the level spacing affects the resonant decay.
The effect of the level spacing on the KAs in NRs
will be reported elsewhere.

\section{Raman intensity}\label{sec:ri}

\subsection{Raman Activity}

In the Raman process, an incident photon 
excites an electron in the valence energy band
into a state in the conduction energy band. 
Then the photo-excited electron emits or absorbs a phonon.
The matrix element for the emission or absorption of a phonon
is given by the el-ph matrix element for the scattering 
between an electron state in the conduction energy band 
and a state in the conduction energy band, 
which is in contrast to that for the el-ph matrix element 
for electron-hole pair creation which is relevant to the matrix element 
from a state in the valence energy band to a state in the conduction
energy band. 
This matrix element for the emission or absorption of a phonon
is given by removing the minus sign 
from $-\phi^J_{\rm B}$ (or $-\phi^I_{\rm uB,dB}$) 
of the final state in 
the electron-hole pair creation matrix elements 
in Eqs.~(\ref{eq:m1}), (\ref{eq:m2}), and (\ref{eq:U_arm}).
This operation is equivalent to  
replacing ${\rm Im}$ (${\rm Re}$) with ${\rm Re}$ (${\rm Im}$)
in Eqs.~(\ref{eq:a}), (\ref{eq:U_zig}), (\ref{eq:Vud}) and 
(\ref{eq:A_L}).

As a result, 
the Raman intensity of the TO (LO) modes in Z-NRs
is proportional to 
${\rm Re}\left[ \phi^{J*}_{\rm A} \phi^J_{\rm B} \right]$
(${\rm Im}\left[ \phi^{J*}_{\rm A} \phi^J_{\rm B} \right]$).
Thus the TO modes are Raman active, 
while the LO modes are not.
Because the TO modes in Z-NRs are free from the KA,
the G band Raman spectra exhibit
the original frequency of the TO modes, $\hbar \omega_0$.
For A-NRs, on the other hand,
the cancellation between $V_{\rm u}$ and $V_{\rm d}$ 
occurs for the TO modes, and 
the TO modes are then not Raman active, while 
the LO modes are Raman active.
Since the LO modes in A-NRs undergo KAs,
the renormalized frequencies, $\hbar \omega_0 + \Pi(\omega_0)$, 
will appear below the original frequencies of the LO modes
by about 20 cm$^{-1}$ (see Fig.~\ref{fig:EF}). 
Thus, the G band spectra in A-NRs can appear 
below those in Z-NRs, which 
is useful in identifying the orientation of the edge of NRs
by G band Raman spectroscopy. 
Our results are summarized in TABLE~\ref{tab:1}
combined with the results for armchair and zigzag SWNTs.

For the Raman intensity of armchair SWNTs, 
we obtain the same conclusion as that of Z-NRs, that is,
the TO modes are Raman active, while the LO modes are not.
For metallic zigzag SWNTs, on the other hand,
the cancellation between $V_{\rm u}$ and $V_{\rm d}$ 
which occurs for the TO modes in A-NRs is not valid. 
Then the TO modes, in addition to the LO modes,
can be Raman active.
However, as we will show in Sec.~\ref{sec:ps}, 
since the matrix element for the emission or absorption of the TO modes 
vanishes at the van Hove singularities of the electronic sub-bands, 
then we can conclude that the TO modes are hardly excited
in resonant Raman spectroscopy.
It is interesting to compare these results
with another theoretical results for the Raman intensities of SWNTs.
In Ref.~\onlinecite{saito01}, 
it is shown using bond polarization theory that 
the Raman intensity is chirality dependent.
In particular, for an armchair (zigzag) SWNT, 
the $A_{1g}$ TO (LO) mode is a Raman active mode, while 
the $A_{1g}$ LO (TO) mode is not Raman active.
These results for SWNTs are consistent with our results.
We think that it is natural that 
the Raman intensity does not change by unrolling the SWNT
since the Raman intensity is proportional to the number 
of carbon atoms in the unit cell
and is not sensitive to the small fraction of carbon atoms
at the boundary.

\begin{table}[htbp]
 \caption{\label{tab:1} Dependences of the KAs and Raman
 intensities on the $\Gamma$ point optical phonon modes in NRs and
 metallic SWNTs.
 $\bigcirc$ and $\times$ represent `occurrence' and `absence',
 respectively. 
 $\triangle$ means that the KA is possible,
 but the broadening effect weakens due to the presence of the
 edge states.
 $\bigtriangledown$ means that the KA is possible,
 but the $E_{\rm F}$ dependence is suppressed by the decoupling from the
 metallic energy band crossing at the Dirac point.
 }
 \begin{ruledtabular}
  \begin{tabular}{c|ccc}
   & mode & {\bf Kohn anomaly} & {\bf Raman active} \\
   \hline 
   Z-NRs & LO & $\triangle$ & $\times$  \\
         & TO & $\times$ & $\bigcirc$ \\ 
   \hline
   A-NRs & LO & $\bigcirc$ & $\bigcirc$ \\
         & TO & $\times$ & $\times$ \\
   \hline 
   Armchair SWNTs & LO & $\bigcirc$ & $\times$  \\
   (rolled Z-NRs) & TO & $\bigtriangledown$ & $\bigcirc$ \\ 
   \hline
   Zigzag SWNTs & LO & $\bigcirc$ & $\bigcirc$ \\
   (rolled A-NRs) & TO & $\bigtriangledown$ & $\times$ \\
  \end{tabular}
 \end{ruledtabular}
\end{table}

\subsection{Comparison With Experiment}\label{sec:exp}

\begin{figure}[htbp]
 \begin{center}
  \includegraphics[scale=1.0]{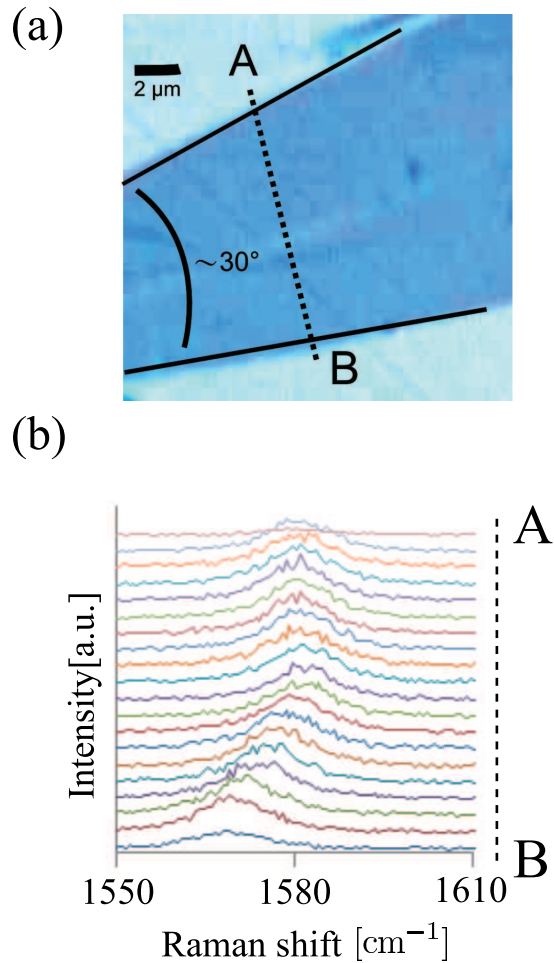}
 \end{center}
 \caption{(color online)
 (a) An optical image of a graphene sample.
 The sample is characterized as a multi-layer graphene 
 ($\sim$ 5 layers) sample.
 (b) The position dependence of the G band frequency.
 From the theoretical results obtained in this paper, 
 we conclude that the G band appearing near the upper (lower) side of
 the edge consists only of the TO (LO) mode, and that 
 the upper (lower) edge
 is dominated by the zigzag (armchair) edge.
 }
 \label{fig:exp}
\end{figure}

We prepare graphene samples by means of the cleavage method 
to observe the frequency of the G band Raman spectra.
In many cases, graphene samples obtained by the cleavage method
show that the angles between the edges have an average value 
equal to multiples of $30^\circ$. 
This is consistent with the results by You {\it et al}.~\cite{you08}
Figure~\ref{fig:exp}(a) shows an optical image of 
the exfoliated graphene with the edges 
crossing each other with an angle of $\sim 30^\circ$.
This angle can be considered as evidence of the presence 
of edges composed predominantly of zigzag or armchair edges. 
Note that, the obtained sample is characterized as a multi-layer
graphene. We estimated the number of layers to be about five 
based on the behavior of the G' band. 
The sample is placed on a SiO$_2$ (100) surface 300 nm in thickness.

The Raman study was performed using a Jobin-Yvon T64000 Raman system.
The laser energy is 2.41eV (514.5nm), the laser power is below 1mW and 
the laser spot is about 1 $\mu m$ in diameter.
Figure~\ref{fig:exp}(b) shows spectra for 
finding the position dependence of the G band.
The results show that the G band frequency 
depends on the position of laser spot. 
When the spot is focused near the upper edge (A)
or far from the edge, the position of the G band 
is almost similar to that of graphite (1582cm$^{-1}$).
However, a softening of the G band is clearly seen 
when the laser spot moves to 
the vicinity of the lower edge (B).

%%%%%%%%%%%%%%%%%%%%%%%%%%%%%
%%%%%%%%%%%%%%%%%%%%%%%%%%%%%

Based on our theoretical studies, we find that 
the G band observed near the upper edge consists only of the TO mode,
while the G band observed near the lower edge comes from the LO mode,
because the G band near the lower edge shows a down shift 
which is considered to be due to the KA effect of the LO mode.
Then, we speculate that the upper (lower) edge is dominated by the
zigzag (armchair) edge.

It should be noted that 
our experiment does not prove that
the observed down shift of the G band near the lower edge
is due to the KA effect, since we do not examine the $E_{\rm F}$
dependence. 
There is a possibility that the observed downshift of the G band is
related to mechanical effects. 
Mohiuddin {\it et al}.~\cite{mohiuddin09} observed that 
the G band splits into two peaks due to uniaxial strain and 
both peaks exhibit redshifts with increasing strain. 
The edge in this work somehow has half of it suspended and this may
decrease the vibration energy.
This effect may explain why for the upper edge in Fig.~\ref{fig:exp} 
the frequencies are slightly downshifted compared with the spectra taken
at the center of the graphene sample.
Moreover, it is probable that the physical edge in this work is a
mixture of armchair and zigzag edges.~\cite{casiraghi09}
Note also that we measured the D band in order to confirm that the
identification of the orientation of the edge is consistent with the fact
that the D band intensity is stronger near the armchair edge than the
zigzag edge.~\cite{pimenta07,you08,malard09}
However, we could not resolve the difference of the intensity 
near the upper and lower edges clearly. 
Zhou et al.~\cite{zhou08} observed by means of high-resolution
angle-resolved photo-emission spectroscopy experiment for epitaxial
graphene that  
the D band gives rise to a kink structure in the electron
self-energy and pointed out that an interplay between the el-ph and
electron-electron interactions plays an important role in the physics 
relating to the D band.
We will consider these issues further in the future.

\section{Pseudospin}\label{sec:ps}

In the preceding sections
we have shown for Z-NRs 
that the LO modes are not Raman active 
and that the TO modes do not undergo KAs.
These results originate from the fact that
the Bloch function is a real number.
Besides, the TO modes in A-NRs are not Raman active and 
do not undergo KAs.
This is due to the cancellation between
$V_{\rm u}$ and $V_{\rm d}$ for the TO modes.
The LO modes in A-NRs undergo KAs
since the Bloch function is a complex number.
The absence or presence of a relative phase 
between the $\phi_{\rm A}$ and $\phi_{\rm B}$
of the Bloch function is essential 
in deriving our results.
In this section, we explain 
the phase of the Bloch function in terms of the
pseudospin,~\cite{sasaki08ptps}  
and clarify the effect of the zigzag and armchair edges 
on the phase of the Bloch function.

\subsection{Absence of a Pseudospin Phase in Z-NRs}

\begin{figure}[htbp]
 \begin{center}
  \includegraphics[scale=0.5]{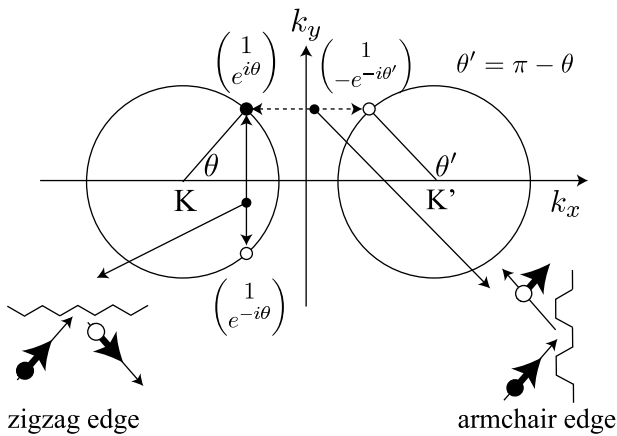}
 \end{center}
 \caption{In $k$-space, 
 we consider a state near the K point 
 (solid circle) and states which are scattered by 
 the zigzag and armchair edges (empty circles). 
 $k_x$ ($k_y$) is taken 
 as parallel to the zigzag (armchair) edge.
 The arrows in the insets indicate 
 the direction of the pseudospins.
 The $\langle \sigma_y \rangle$ component of the pseudospin 
 is reversed at the zigzag edge, while
 the $\langle \sigma_x \rangle$ component 
 is preserved at the armchair edge.
 }
 \label{fig:phase}
\end{figure}

Here we use the effective-mass model
in order to understand the reason why 
the Bloch function in Z-NRs is a real number.
In the effective-mass model, 
the Bloch functions in the conduction energy band 
near the K and K' points are given by~\cite{sasaki08ptps}
\begin{align}
 \begin{split}
  & \varphi_{\rm K}(k_x,k_y)=\frac{1}{\sqrt{2}}
  \begin{pmatrix}
   1 \cr e^{i\theta}
  \end{pmatrix}, \\
  & \varphi_{\rm K'}(k_x,k_y)=\frac{1}{\sqrt{2}}
  \begin{pmatrix}
   1 \cr -e^{-i\theta'}
  \end{pmatrix},
 \end{split}
 \label{eq:kpw}
\end{align}
where $k_x$ and $k_y$ ($k'_x$ and $k'_y$)
are measured from the K (K') point and 
the angle $\theta$ ($\theta'$) is defined by 
$k_x + i k_y \equiv |{\bf k}|e^{i\theta}$
($k'_x + i k'_y \equiv |{\bf k'}|e^{i\theta'}$).
It is noted that $k_x$ ($k_y$) is taken as
parallel to the zigzag (armchair) edge.
Then
$k_y$ is reflected into $-k_y$ at the zigzag edge, 
and the scattered state is given by 
\begin{align}
 \varphi_{\rm K}(k_x,-k_y)=\frac{1}{\sqrt{2}}
 \begin{pmatrix}
  1 \cr e^{-i\theta}
 \end{pmatrix},
 \label{eq:negky}
\end{align}
as shown in Fig.~\ref{fig:phase}.

%%%%%%%%%%%%%%%%%%%%%%%%%%%%%
%%%%%%%%%%%%%%%%%%%%%%%%%%%%%

The relative phase of the wave function at the A and B sublattices 
can be characterized by the direction of the pseudospin.
The pseudospin is defined by the expectation value 
of the Pauli matrices $\sigma_{x,y,z}$ 
with respect to the Bloch function.~\cite{sasaki08ptps}  
For $\varphi_{\rm K}(k_x,k_y)$,
we have $\langle \sigma_x \rangle = \cos \theta$,
$\langle \sigma_y \rangle = \sin \theta$,
$\langle \sigma_z \rangle = 0$.
For $\varphi_{\rm K}(k_x,-k_y)$,
we have $\langle \sigma_x \rangle = \cos \theta$,
$\langle \sigma_y \rangle = -\sin \theta$,
$\langle \sigma_z \rangle = 0$.
Then $\langle \sigma_y \rangle$
flips at the zigzag edge as shown in Fig.~\ref{fig:phase}.
Thus, due to an interference between the incoming and reflected waves,
we have $\langle \sigma_y \rangle=0$ 
for the Bloch function near the zigzag edge.
The condition $\langle \sigma_y \rangle=0$
means that the Bloch function becomes a real number. 
In fact, since 
$\langle \sigma_y \rangle=2 \sum_J {\rm Im}\left[ \phi_{\rm A}^{J*}
\phi_{\rm B}^{J} \right]$ and 
$\langle \sigma_x \rangle=2 \sum_J {\rm Re}\left[ \phi_{\rm A}^{J*}
\phi_{\rm B}^{J} \right]$ for the Bloch function 
$(\phi_{\rm A}^J,\phi_{\rm B}^J)$ 
in Eq.~(\ref{eq:wfz}), 
the el-ph matrix elements $V_{\rm T}$ and $V_{\rm L}$
(Eq.~(\ref{eq:a})) are proportional to 
$\langle \sigma_y \rangle$ and $\langle \sigma_x \rangle$,
respectively. 
In Appendix~\ref{app:rwf}, we show the relationship
between the Bloch function $\varphi_{\rm K}$ 
in the effective-mass model (Eq.~(\ref{eq:kpw})) 
and the Bloch function $(\phi_{\rm A}^J,\phi_{\rm B}^J)$ 
in the tight-binding model (Eq.~(\ref{eq:wfz})).

We point out that 
the condition $\langle \sigma_y \rangle=0$
is not satisfied in the case of armchair SWNTs (rolled Z-NRs)
except for the lowest energy sub-bands of $k_y=0$.
%When $k_y\ne 0$, we can make a symmetric state
%($\phi_+$) and an antisymmetric state ($\phi_-$)
%from two degenerate states with $(k_x,k_y)$ and $(k_x,-k_y)$ as 
%\begin{align}
% \phi_\pm
% = e^{ik_y y} \varphi_{\rm K}(k_x,k_y) \pm 
% e^{-ik_y y} \varphi_{\rm K}(k_x,-k_y).
%\end{align}
%Since $\phi_\pm$ is also an energy eigenstate 
%as well as $\varphi_{\rm K}(k_x,k_y)$ and $\varphi_{\rm K}(k_x,-k_y)$, 
%we can use it as a basis for the Bloch function. 
%Because of Eq.~(\ref{eq:negky}), 
%$\phi^\pm_{\rm A}$ and $\phi^\pm_{\rm B}$ 
%do not have a relative phase and we obtain 
%$\langle \sigma_y \rangle=0$ for $\phi_\pm$.
This is the reason why we see in Fig.~\ref{fig:EF}(a) 
that the TO mode exhibits no broadening but a softening 
with a constant energy $\sim -30$ cm$^{-1}$ in a $(5,5)$ SWNT.
It is interesting to note that 
the Aharanov-Bohm flux applied along the tube axis
shifts the cutting lines~\cite{ajiki93,zaric04,minot04} 
and can make $k_y$ for the lowest energy sub-bands nonzero.
Thus the Aharanov-Bohm flux makes that
$\langle \sigma_y \rangle=\sin \theta \ne 0$
even for the lowest energy sub-bands, for which 
the TO mode can exhibit a broadening.~\cite{ishikawa06}

Since the effective-mass model describes the physics well
in the long wave length limit, 
an advantage in the above discussion of using the effective-mass model 
is that it is not necessary for the edge of graphene NRs 
to be well defined on an atomic scale in order that 
we have a cancellation of $\langle \sigma_y \rangle$.
This may be a reason why we observe a softening of the G band
near the edge of an armchair-rich sample
experimentally as shown in Sec.~\ref{sec:exp}.

\subsection{Coherence of the Pseudospin in A-NRs}

On the other hand,
a state near the K point with $(k_x,k_y)$ 
is reflected by the armchair edge
into a state near the K' point 
with $(k'_x,k'_y)=(-k_x,k_y)$,
and the scattered state is given by $\varphi_{\rm K'}(-k_x,k_y)$.
In this case, by putting $\theta'=\pi-\theta$
into $\varphi_{\rm K'}(k_x,k_y)$ in Eq.~(\ref{eq:kpw}), 
we obtain 
\begin{align}
 \varphi_{\rm K'}(-k_x,k_y)
 =\frac{1}{\sqrt{2}}
  \begin{pmatrix}
   1 \cr e^{i\theta}
  \end{pmatrix}
\end{align}
which is identical to the initial Bloch function, 
$\varphi_{\rm K}(k_x,k_y)$.
Thus the relative phase between
the A and B Bloch functions
is conserved thorough the reflection by the armchair edge
so that the Bloch function can not be reduced to a real number.
Namely, the reflections by the armchair edge 
preserve the pseudospin as shown in Fig.~\ref{fig:phase}.
It is expected that 
the relative phase makes it possible that
KAs occur not only for the LO mode but also for the TO mode 
near the armchair edge.
However, as we have shown in Eq.~(\ref{eq:cor}), 
the armchair edge gives rise to the cancellation
between $V_{\rm u}$ and $V_{\rm d}$, so that 
the TO modes in A-NRs do not undergo KAs.
We consider whether Eq.~(\ref{eq:cor}) is satisfied 
in the case of zigzag SWNTs or not, 
in order to see if the KA effect is present in zigzag SWNTs
or not. 
By applying the Bloch theorem to zigzag SWNTs, we have
\begin{align}
\begin{split}
 & \phi_{\rm uA}^{I}=(e^{iIka}/2\sqrt{N}) \varphi_{\rm A}, \\
 & \phi_{\rm uB}^{I}=(e^{i(I+1/2)ka}/2\sqrt{N}) \varphi_{\rm B}, \\
 & \phi_{\rm dA}^{I}=(e^{i(I+1/2)ka}/2\sqrt{N}) \varphi_{\rm A}, \\
 & \phi_{\rm dB}^{I}=(e^{iIka}/2\sqrt{N}) \varphi_{\rm B},
\end{split}
\end{align}
where we set $\varphi_{\phi} =
{}^t (\varphi_{\rm A},\varphi_{\rm B})$.
Using these equations, we obtain 
\begin{align}
\begin{split}
 & \sum_I \phi_{\rm uA}^{I*} \phi_{\rm uB}^I
 = \frac{1}{4}e^{i\frac{ka}{2}} 
 \varphi_{\rm A}^* \varphi_{\rm B},
 \\
 & \sum_I \phi_{\rm dA}^{I*} \phi_{\rm dB}^I 
 = \frac{1}{4}e^{-i\frac{ka}{2}} 
 \varphi_{\rm A}^* \varphi_{\rm B}.
\end{split}
 \label{eq:ziga}
\end{align}
Thus the first equation in Eq.~(\ref{eq:cor})
is not satisfied for zigzag SWNTs.
Similarly, we have 
\begin{align}
\begin{split}
 & \sum_I \phi_{\rm uA}^{I+1*} \phi_{\rm uB}^I =
 \frac{1}{4}e^{-i\frac{ka}{2}} 
 \varphi_{\rm A}^* \varphi_{\rm B},
 \\
 & \sum_I \phi_{\rm dA}^{I*} \phi_{\rm dB}^{I+1}
 = \frac{1}{4}e^{i\frac{ka}{2}} 
 \varphi_{\rm A}^* \varphi_{\rm B},
\end{split}
\label{eq:zigb}
\end{align}
which shows that the second equation in Eq.~(\ref{eq:cor})
is not fulfilled, either.
Thus the TO modes 
can undergo KAs because 
the cancellation between 
$V_{\rm u}$ and $V_{\rm d}$ is not possible
for zigzag SWNTs.
In fact, 
by putting Eqs.~(\ref{eq:ziga}) and (\ref{eq:zigb}) 
into Eq.~(\ref{eq:Vud}), we get
\begin{align}
 V_{\rm u}+V_{\rm d} &= 
 i (\delta \gamma_{0,2} + \delta \gamma_{0,3}) \sin \theta \cos
 \frac{ka}{2} \nn \\
 & + i(\delta \gamma_{0,3} - \delta \gamma_{0,2}) \cos \theta \sin
 \frac{ka}{2},
\end{align}
where we set $\varphi_{\rm A}^* \varphi_{\rm B}=e^{i\theta}$.
Because the TO modes satisfy $\delta \gamma_{0,2} + \delta \gamma_{0,3}=0$,
$V_{\rm u}+V_{\rm d}$ can take a nonzero value for 
\begin{align}
 V_{\rm u}+V_{\rm d} = i 2 \delta \gamma_{0,3}
 \sin\left( \frac{ka}{2} \right) \cos \theta.
 \label{eq:Vud_zswnt}
\end{align}

It is noted that Eq.~(\ref{eq:Vud_zswnt}) vanishes 
when $\theta =\pm \pi/2$.
This condition $\theta =\pm \pi/2$ 
is satisfied for 
low energy electron-hole pairs 
when the energy band crosses the Dirac point.
In other words, 
high energy electron-hole pairs in the sub-bands 
can contribute to a frequency softening of the TO mode
in zigzag SWNTs. 
This is why we obtain the down shift of the TO mode
for a $(9,0)$ zigzag SWNT as shown in Fig.~\ref{fig:EF}.
In ``metallic'' zigzag SWNTs,
the curvature effect shifts 
the position of the cutting line~\cite{saito:153413}
of the metallic energy band
from the Dirac point
and produces a small energy gap.~\cite{saito92prb,kane97,ando00,yang00}
In this case, 
the low energy electron-hole pairs satisfy
$\cos \theta \ne 0$ in Eq.~(\ref{eq:Vud_zswnt})
and they contribute to a frequency hardening of the TO
modes in metallic zigzag SWNTs.
The curvature effect gives rise to a change of the Fermi surface
and results in KAs for the TO modes.~\cite{sasaki08_curvat}

Similarly, 
the matrix element for the emission or absorption of the TO modes
in zigzag SWNTs is given by
\begin{align}
 2 \delta \gamma_{0,3}
 \sin\left( \frac{ka}{2} \right) \sin \theta, 
\end{align}
which does not vanish in general.
This shows that the TO modes in zigzag SWNTs can be Raman active. 
However, since the van Hove singularities of sub-bands in zigzag SWNTs
are located on the $k_x$ axis (and satisfy $\theta=0$), 
the $\sin \theta$ factor tells us that
the TO modes are hardly excited in resonant Raman spectroscopy.

\section{Discussion}\label{sec:dis}

The theoretical analysis performed in this paper is based on 
the use of a simple tight-binding method 
which includes only the first nearest-neighbor hopping integral 
and its variation due to the atomic displacements. 
The approximation used is partly justified because 
the deformation-potential and the el-ph matrix element with
respect to the second nearest-neighbors is about one order of magnitude
smaller than that of the first nearest neighbors.~\cite{porezag95,jiang05prb}
However, we have not considered the effect of the overlap integral. 
The overlap integral breaks the particle-hole symmetry 
and may invalidate our results. 
Besides, we have neglected the contribution of 
momentum non-conserving electron-hole pair creation processes 
in evaluating $\sum_{\rm eh}$ in Eq.~(\ref{eq:PI}).
Although this is an approximation which works well for thin NRs,
the inclusion of the momentum non-conserving electron-hole pair creation
processes may invalidate our results.
We will elaborate on this idea in the future.

\section{Summary}\label{sec:sum}

In summary, 
the LO modes undergo KAs in graphene NRs 
while the TO modes do not.
This conclusion does not depend on the orientation of the edge.
In Z-NRs, the Raman intensities of the LO modes 
are strongly suppressed because the wave function is a real number, 
and only the TO modes are Raman active.
As a result, the KA for the LO mode in Z-NRs
would be difficult to observe in Raman spectroscopy. 
In A-NRs, only the LO modes are Raman active 
owing to the cancellation between $V_{\rm u}$ and $V_{\rm d}$.
The ``chirality'' dependent Raman intensity derived for NRs
is the same as the chirality dependent Raman intensity for SWNTs 
calculated in Ref.~\onlinecite{saito01}.
The strong down shift of the LO mode makes it possible to 
identify the orientation of edges of graphene by the G band Raman spectroscopy
due to the ``chirality'' dependent Raman intensity.

\section*{Acknowledgment}

K.S would like to thank Hootan Farhat and Prof. Jing Kong
for discussions on the KAs in SWNTs. 
S.M acknowledges MEXT Grants (No.~21000004 and No.~19740177).
R.S acknowledges a MEXT Grant (No.~20241023).
M.S.D acknowledges grant NSF/DMR 07-04197.
This work is supported by 
a Grant-in-Aid for Specially Promoted Research
(No.~20001006) from MEXT.

\appendix

\section{Derivation of Eq.~(\ref{eq:cor})}\label{app:mirror}

In this Appendix, we derive Eq.~(\ref{eq:cor}) 
by using mirror and time-reversal symmetries. 
Let us introduce the mirror-reflection operator $M$ by
\begin{align}
 M \phi^I_{n,k} =
 \begin{pmatrix}
  0 & 0 & 0 & 1 \cr
  0 & 0 & 1 & 0 \cr
  0 & 1 & 0 & 0 \cr
  1 & 0 & 0 & 0
 \end{pmatrix}
 \begin{pmatrix}
  \phi^I_{\rm uA} \cr
  \phi^I_{\rm uB} \cr
  \phi^I_{\rm dA} \cr
  \phi^I_{\rm dB}
 \end{pmatrix}
 \ \ (I=0,\ldots,N),
\end{align}
where $k$ is the wave vector along the armchair edge
and $n$ is the band index. 
By applying $M$ to the energy eigen-equation
$H_k \phi_{n,k}=E_{n,k} \phi_{n,k}$,
we get
$M H_k \phi_{n,k}=E_{n,k} M \phi_{n,k}$.
Since the Hamiltonian satisfies $M H_k M^{-1} =H_{-k}$,
we obtain $M \phi_{n,k} = e^{i\phi} \phi_{n,-k}$
where $\phi$ is a phase factor.

On the other hand, 
due to the time-reversal symmetry, 
we have $\phi_{n,k}^* = e^{i\phi'} \phi_{n,-k}$.
Thus, by combing the time-reversal symmetry ($\phi_{n,k}^* =e^{i\phi'} \phi_{n,-k}$)
with the mirror symmetry ($M \phi_{n,k} = e^{i\phi}\phi_{n,-k}$),
we get
\begin{align}
 M \phi_{n,k} =e^{i\phi''}\phi_{n,k}^*,
\end{align}
that is, 
\begin{align}
 \begin{pmatrix}
  \phi^I_{\rm dB} \cr
  \phi^I_{\rm dA} \cr
  \phi^I_{\rm uB} \cr
  \phi^I_{\rm uA}
 \end{pmatrix}
 =e^{i\phi''}
 \begin{pmatrix}
  \phi^{I*}_{\rm uA} \cr
  \phi^{I*}_{\rm uB} \cr
  \phi^{I*}_{\rm dA} \cr
  \phi^{I*}_{\rm dB}
 \end{pmatrix}.
\end{align}
Using this condition, 
one sees that Eq.~(\ref{eq:cor}) is satisfied.

\section{Relationship between 
Eq.~(\ref{eq:wfz}) and Eq.~(\ref{eq:kpw})}\label{app:rwf}

Here we will show for Z-NRs that 
the Bloch function derived using the tight-binding lattice model
(Eq.~(\ref{eq:wfz})) is a superposition of 
incoming and reflected Bloch functions derived using the effective-mass
model (Eq.~(\ref{eq:kpw})).

By rewriting the Bloch function of Z-NRs
$(\phi_{\rm A}^J,\phi_{\rm B}^J)$ in Eq.~(\ref{eq:wfz}) as
\begin{align}
 \begin{split}
  & \phi_{\rm A}^J = 
  \frac{C(g,\phi)}{2i}
  \left( \frac{1}{g} + e^{i\phi} \right) e^{iJ\phi} 
  +{\rm c.c.}, \\
  & \phi_{\rm B}^J = 
  \frac{C(g,\phi)}{2i} \epsilon(g,\phi) \frac{e^{i\phi}}{g} 
  e^{iJ\phi} + {\rm c.c.},
 \end{split}
 \label{eq:wfz-loc}
\end{align}
where ${\rm c.c.}$ denotes the complex conjugation
of the first term,
one sees that 
$(\phi_{\rm A}^J,\phi_{\rm B}^J)$ is a real number
as a result of the cancellation of the imaginary part
between the first and second terms.
By introducing a new Bloch function $\varphi_\phi$ as
\begin{align}
 \varphi_{\phi} \equiv 
 \frac{C(g,\phi)}{2i g}e^{i\phi}
 \begin{pmatrix}
  g + e^{-i\phi} \cr
  \epsilon(g,\phi)
 \end{pmatrix},
 \label{eq:varphi}
\end{align}
the Bloch function $(\phi_{\rm A}^J,\phi_{\rm B}^J)$ 
is expressed by 
\begin{align}
 \begin{pmatrix}
  \phi_{\rm A}^J \cr \phi_{\rm B}^J
 \end{pmatrix}
 = \varphi_{\phi} e^{iJ\phi} +
 \varphi_{\phi}^* e^{-iJ\phi}.
 \label{eq:wfz-bl}
\end{align}
Because of the different signs in the exponents of 
$e^{iJ\phi}$ and $e^{-iJ\phi}$ in Eq.~(\ref{eq:wfz-bl}),
$\phi$ may be thought of as 
the wave vector perpendicular to the zigzag edge ($k_y$)
multiplied by a lattice constant ($\ell$) as $\phi=k_y \ell$. 
Assuming that $\phi=k_y \ell$,
the zigzag edge reflects 
a state with $k_y$ ($\varphi_{\phi} e^{iJ\phi}$) into 
a state with $-k_y$ ($\varphi_{-\phi} e^{-iJ\phi}$).
We then expect that 
the Bloch function near the zigzag edge 
is given by
\begin{align}
 \varphi_{\phi} e^{iJ\phi} + \varphi_{-\phi} e^{-iJ\phi}.
 \label{eq:blochsum}
\end{align}
It is noted that
Eq. (\ref{eq:blochsum}) is different from Eq.~(\ref{eq:wfz-bl}) 
because $\varphi_{-\phi}$ is not identical to $\varphi_{\phi}^*$ 
in general.
However, we will get $\varphi_{-\phi}= \varphi_{\phi}^*$
for Z-NRs because we may assume that 
the normalization constant $C(g,\phi)$ in Eq.~(\ref{eq:varphi})
satisfies $C(g,-\phi)=-C^*(g,\phi)$ without loss of generality.
Therefore, Eq.~(\ref{eq:blochsum})
is consistent to Eq.~(\ref{eq:wfz-bl}),
which indicates that the assumption ($\phi=k_y \ell$)
is appropriate. 
The condition $\varphi_{-\phi}= \varphi_{\phi}^*$
is a non-trivial condition since 
it is satisfied only for the zigzag edge
and is essential for $(\phi_{\rm A}^J,\phi_{\rm B}^J)$
to be a real number. 

Using Eq.~(\ref{eq:kpw}) we obtain 
$\varphi_{\rm K}(k_x,-k_y)=\varphi_{\rm K}(k_x,k_y)^*$
which corresponds to $\varphi_{-\phi}= \varphi_{\phi}^*$.
In fact, 
by putting $(g+e^{-i\phi})\equiv |g+e^{-i\phi}|e^{-i\Theta}$
into Eq.~(\ref{eq:varphi}) and by using 
$\epsilon(g,\phi) = |g+e^{-i\phi}|$,~\cite{sasaki05prb}
the Bloch state $\varphi_{\phi}$ can be written as
\begin{align}
 \varphi_{\phi}=\frac{1}{\sqrt{2}}
  \begin{pmatrix}
   1 \cr e^{i\Theta}
  \end{pmatrix},
\end{align}
where $\Theta$ and $\theta$ are the same as each other 
near the K point.

%\bibliographystyle{apsrev}
% \bibliography{/home/sasaki/bib/sasaki_mgm,/home/sasaki/bib/sasaki}
%\bibliography{/misc/home1/staff/sasaken/bib/sasaki_mgm,/misc/home1/staff/sasaken/bib/sasaki}
% \bibliography{/rsaito/bib/mgm,/rsaito/bib/gic,/rsaito/bib/mag_gic,/rsaito/bib/c60,/rsaito/bib/carbon,/rsaito/bib/fiber,/rsaito/bib/alex,/home/staff/jiang/bib/jiang,/misc/home1/staff/sasaken/bib/sasaki}

\end{document}